\documentclass[11pt]{article}

\usepackage{amsmath,amssymb,amsthm}
\usepackage{geometry}
\usepackage{bm}
\usepackage{graphicx}

\begin{document}

\title{On the Possibility of Quantum Gravity Emerging from Geometry}

\author{Jaume Gin\'e}
\date{ Departament de Matem\`atica, Universitat de Lleida, \\
Avda. Jaume II, 69; 25001 Lleida, Catalonia, Spain \\
{\small {\rm E--mail:} {\tt jaume.gine@udl.cat}}}

\maketitle

\begin{abstract}
Can the generalized uncertainty principle (GUP) arise as an effective manifestation of spacetime geometry rather than as a fundamental postulate? More specifically, can the gravitational GUP be interpreted as an emergent uncertainty relation generated by the statistical microstructure of local horizons? Motivated by these questions, this work explores the broader possibility that selected features of quantum-gravitational behavior may originate from the geometric and thermodynamic properties of spacetime itself. We argue that, under suitable assumptions regarding the multifractal structure of microscopic horizons, an effective GUP can indeed be induced by geometry. This perspective suggests that quantum uncertainty may be viewed not as a fundamental ingredient of nature, but as an emergent consequence of the statistical organization of spacetime at the Planck scale. While the present framework does not constitute a complete theory of quantum gravity, nor does it derive the full formalism of quantum mechanics, it provides a proof of concept for a geometric origin of generalized uncertainty relations and establishes a thermodynamic route connecting horizon microstructure, entropy corrections, and effective quantum behavior. The resulting framework is compatible with several modern approaches to quantum gravity through the common notion of scale-dependent geometry, while yielding a unified geometric interpretation of generalized uncertainty relations.
\end{abstract}

\section{Introduction}

In this paper, we consider that the spacetime is a random multifractal metric space at Planckian resolution characterized by
a spectrum of local Hausdorff dimensions $D(q)$, and where classical spacetime is recovered when $D(q)\equiv 2$. More specifically, the multifractal spatial geometry is defined as follows.
Let $(\Sigma,\mu)$ be a spatial slice equipped with a probability measure $\mu$ such that for every $x\in\Sigma$, we have
$\mu(B_r(x)) \sim r^{D(q(x))}$, for $r$ tending to zero, where $B_r(x)$ is the metric ball of radius $r$, $D(q)\in(0,3]$ is a local Hausdorff dimension, $q$ labels the multifractal sectors, and $w(q)$ is a normalized weight measure with $\int dq\, w(q)=1$.
Moreover we consider that there exists a minimal geometric resolution $\ell_{P}>0$ such that no physical observable probes lengths below $\ell_{P}$ and all geometric quantities are defined relative to this scale, see \cite{GG}.
We assume, following \cite{Gine}, a scale-dependent effective geometry. Probing geometry at scale $\ell \ge \ell_{P}$ yields an effective area/volume renormalization
\[
\alpha(\ell) := \int dq\, w(q)\left(\frac{\ell}{\ell_{P}}\right)^{2 - D(q)},
\]
that encodes auto-affine and multifractal anisotropy. In this framework we will see that the quantum uncertainty is induced by geometry, i.e.
quantum uncertainty relations are emergent constraints induced by microscopic geometry.
Indeed, we will see that the effective position uncertainty satisfies
\[
\Delta x \gtrsim \frac{\hbar}{2\Delta p} \;+\; \gamma\,\frac{\ell_{P}^{2}}{\hbar}\,\Delta p,
\]
where the geometric coupling is
\[
\gamma := \int dq\, w(q)\,[D(q)-2]^{2}.
\]
that is the variance of geometric dimensional fluctuations.

Hence we generalize uncertainty principle (GUP) as a geometric effect. One can obtain a GUP without any postulating modified commutators and without quantizing gravity,
but instead from microscopic geometric structure of spacetime. In particular, from horizon microgeometry (rough, fractal, multifractal, auto-affine) like is made in \cite{Gine} inspired in \cite{Barrow}. Schematically,
\begin{equation}\label{GUPL}
\Delta x \gtrsim \frac{\hbar}{2\,\Delta p} \;+\; \alpha(\ell)\,\frac{\ell_{P}^{2}}{\hbar}\,\Delta p ,
\end{equation}
where the second term represents geometric backreaction. This implies that the gravitational GUP can be reinterpreted as an \emph{effective kinematical consequence} of spacetime geometry. This is a dynamical reinterpretation, not a philosophical one, i.e., is a consequence of the geometry. If uncertainty relations depend on geometry, then geometry is no longer merely a background, geometry controls the structure of phase space.

This leads to the conceptual leap that quantum behavior may emerge from the \emph{statistical geometry} of spacetime itself.
Next we will see gravity as emergent kinematics. This idea lies at the intersection of emergent gravity, spacetime microstructure, and effective field theory.

\smallskip

Finally we claim that Quantum Gravity (QG) is also emerging from geometry.
The first step is to establish effective Quantum Gravity, and this is achievable now from the results described in \cite{Gine}.
Microscopic geometric irregularities induce effective quantum corrections in the GUP, logarithmic entropy corrections, modified evaporation, and modified dispersion relations. This corresponds to semiclassical gravity and an Effective Field Theory (EFT) on non-smooth backgrounds.

\medskip

The second step is the emergent of quantization (non-commutativity) from geometry.
If uncertainty relations are geometric, then canonical commutators \([x,p]=i\hbar\)
are not fundamental but emergent. One may obtain
\[
[x,p]_{\text{eff}} = i\hbar\left(1 + \alpha p^{2} + \cdots\right),
\]
where $\alpha$ is a parameter that measures the average size of fluctuations of the underlying spacetime geometry.
This connects with the Kempf--Mangano--Mann GUP \cite{Kem}, Snyder geometry \cite{Sny}, and non-commutative spacetime.
But here the origin is geometric rather than algebraic.

\medskip

The final objective is to achieve the full Quantum Gravity theory. A complete theory would require:
a dynamical theory of microscopic geometry, derivation of Hilbert space, superposition, and unitarity, recovery of QFT in flat space and GR at large scales. Up to now, no existing framework fully accomplishes this.

\medskip

The geometric GUP picture naturally aligns with the existing Quantum Gravity programs as
Loop Quantum Gravity \cite{Rov,Ash,AL} (discrete area spectrum, horizon microstates, logarithmic entropy corrections), Asymptotic Safety \cite{Reu,RS} (scale-dependent geometry, running spectral dimension; where $D(q)$ plays a similar statistical role), Causal Dynamical Triangulations \cite{Amb,Amb2,Loll}(emergent 4D geometry, fractal behaviour at Planck scales), and Generalized statistical mechanics \cite{Ka,Tsa} (Kaniadakis, Tsallis) (deformed phase space and modified dispersion relations, with deformation induced by horizon geometry).

Here we are not claiming that geometry replaces quantum mechanics, and that gravity can be quantized without quantization.
What we claim is that quantum behaviour may emerge as an effective description, the GUP is a geometric phenomenon, and that quantum gravity may be a statistical theory of spacetime microgeometry.

The results presented here suggest that generalized uncertainty principles need not be postulated as fundamental quantum deformations, but may instead emerge as effective kinematical relations induced by microscopic spacetime geometry. In this sense, quantum gravitational effects appear as statistical manifestations of horizon microstructure rather than as direct consequences of quantization. While this does not yet constitute a full theory of quantum gravity, it provides evidence that quantum behavior may be deeply rooted in geometric and statistical properties of spacetime itself.

Therefore, from the current results \cite{Gine} and the results proved here, it is possible to develop an effective, semiclassical, and with statistical sense Quantum Gravity emerging from geometry, but not yet a complete fundamental theory, although the framework provides a meaningful step toward that goal.

\section{Emergent GUP from Horizon Geometry}

Let spacetime possess a microscopic horizon geometry characterized by an ensemble of local scaling exponents $D(q)$ with weights $w(q)$.
Then the operational uncertainty in position and momentum measurements performed near the horizon satisfies an effective generalized uncertainty principle
\begin{equation}\label{eq1}
\Delta x \gtrsim \frac{\hbar}{2\,\Delta p} \;+\; \gamma\,\frac{\ell_{P}^{2}}{\hbar}\,\Delta p, \quad  \mbox{where} \quad \gamma \sim \int dq\, w(q)\,\big(D(q)-2\big)^{2},
\end{equation}
where the first term is purely quantum and the second term is purely geometric and it arises from multifractal horizon fluctuations, it is independent of quantization assumptions, and the Planck length enters as the UV resolution scale, not as an imposed cutoff.
Hence, the gravitational GUP emerges as a kinematical consequence of horizon microgeometry. In Eq. (\ref{eq1}) $\gamma$ is a dimensionless variance of the multifractal spectrum around the classical value $D=2$ and it measures the roughness/irregularity strength of the horizon but scale-independent. We will see that this $\gamma$ is the correct object if you want a GUP parameter analogous to $\beta$ in the standard gravitational GUP. However, in \cite{Gine} a generalization of Eq. (\ref{eq1}) was obtained from the scale-dependent factor $\alpha(\ell)$ given by
\[
\alpha(\ell) \sim \int dq\, w(q)\left(\frac{\ell}{\ell_{P}}\right)^{2-D(q)}.
\]
This $\alpha(\ell)$ is a scale-dependent geometric factor that describes how horizon area or entropy scales when probed at resolution $\ell$ and is not a coupling constant, but it is a running function. Indeed, $\alpha(\ell)$ appears in scale-dependent entropy corrections where $S(\ell) \sim \alpha(\ell) A /(4\ell_{P}^2)$, in evaporation rates, and describes a scale-dependent near-horizon physics.

The coefficients $\gamma$ and $\alpha(\ell)$ are connected by a small-$\Delta$ expansion. In the sense that taking
$D(q) = 2 + \Delta(q)$, with $|\Delta(q)| \ll 1$ we have
\[
\left(\frac{\ell}{\ell_{P}}\right)^{2-D(q)}
= \exp\!\left[-\Delta(q)\,\ln\!\left(\frac{\ell}{\ell_{P}}\right)\right],
\]
and expanding
\[
\alpha(\ell) =1 - \langle \Delta(q) \rangle \ln\!\left(\frac{\ell}{\ell_{P}}\right) + \frac{1}{2}\langle \Delta^{2}(q) \rangle
\ln^{2}\!\left(\frac{\ell}{\ell_{P}}\right) + \cdots.
\]
Here, the first moment $\langle \Delta(q) \rangle$ shifts entropy linearly and the second moment $\langle \Delta^{2}(q) \rangle$ controls fluctuations. The first moment can be absorbed into a redefinition of the classical area term. i.e., it rescales the area law (entropy normalization)
especially if you evaluate near a fixed reference scale (e.g. $\ell \sim \ell_P$), where $\ln(\ell/\ell_P)$ is order one. Moreover, it is not universal because depends on the sign of $\Delta(q)$. Hence, the significant contribution is of the second moment and $\gamma$ is
\[
\gamma \propto \langle \Delta^{2}(q) \rangle = \int dq\, w(q)\,\big(D(q)-2\big)^{2},
\]
taking into account that near--Planck--scale approximation we have $\ell \sim \Delta x \sim 1/\Delta p$ and the GUP is evaluated close to the minimal length, where $\Delta p \sim 1/\ell_{P}$, and then the ratio $\ell/\ell_{P}$ is of order unity. Consequently,
$\ln (\ell/\ell_{P}) \sim \mathcal{O}(1)$ and $\ln^2 (\ell/\ell_{P}) \sim \mathcal{O}(1)$ and these logarithmic factors may be absorbed into effective constants (for example into the parameter $\gamma$).
Hence from the multifractal GUP (\ref{GUPL}) we arrive to (\ref{eq1}) using the near--Planck--scale approximation. That is, the logarithmic multifractal correction to the effective resolution, when expressed in terms of momentum uncertainty and expanded near the Planck scale, reduces to a polynomial correction proportional to $(\ell_{P}^{2}/\hbar)\,\Delta p$, yielding the generalized uncertainty principle (\ref{eq1}).

\smallskip

Therefore, the GUP coefficient  comes from the quadratic fluctuation of the multifractal spectrum, not from the full scaling factor. Hence, the generalized uncertainty principle, is controlled by the scale-independent variance of the multifractal spectrum, which measures the intrinsic roughness of spacetime geometry. This $\gamma$ is universal, independent of the probe scale $\ell$, and fixed once geometry is fixed. Otherwise the uncertainty principle would depend on the measurement scale. In this case the minimal length or the minimal position uncertainty is
$(\Delta x)_{\min} = \sqrt{\gamma}\,\ell_P$, which vanishes if and only if $D(q)\equiv 2$.
An equivalent form to arrive to (\ref{eq1}) is to use the effective spatial resolution induced by geometry given by
\[
\ell_{\mathrm{eff}}(\ell)
= \ell_P\left( \int dq\, w(q)\left(\frac{\ell}{\ell_P}\right)^{2-D(q)} \right)^{1/2}.
\]
and doing the same approximations we also arrive to (\ref{eq1}).

\smallskip

The standard gravitationally GUP motivated by the works of \cite{Med,AS,ACS,Kem} is
\[
\Delta x \gtrsim \frac{\hbar}{2\Delta p} \;+\; \beta\,\frac{G}{c^{3}}\,\Delta p ,
\]
can now be reinterpreted as $\beta \equiv \gamma \sim \langle \Delta^{2}(q) \rangle $. More specifically, the previous results we obtain a geometric origin of standard gravitational GUP and parameter $\beta$ satisfies $\beta \sim \gamma$, that is the variance of the multifractal dimension, not an independent constant. Moreover, any quantum system probing spacetime geometry at Planckian scales obeys the same GUP, independently of its microscopic dynamics.

Therefore, gravity does not modify quantum mechanics and is the geometry that modifies the phase space available to quantum excitations.
Hence, we remove the need to postulate modified commutators, introduce non-locality by hand, and invoke speculative Planck-scale dynamics.

The non-commutativity also emerges from geometry. From the geometric GUP (\ref{eq1}), one may define an effective commutator:
\begin{equation}\label{eq2}
[x,p]_{\mathrm{eff}}=i\hbar\left(1 + \gamma\,\frac{p^{2}}{m_{P}^{2}} + \cdots\right).
\end{equation}
where this commutator is not postulated, is scale-dependent, and emerges after coarse-graining over horizon microstructure.
This reproduces in a natural way the Snyder \cite{Sny} noncommutative geometry, the Kempf--Mangano--Mann \cite{Kem} deformation, and Doubly Special Relativity(DSR)-inspired phase spaces \cite{AC,MS}, but with a geometric origin.

\smallskip

The multifractal structure encoded in $w(q)$ and $D(q)$ is postulated rather than derived, and the key geometric coupling $\gamma \sim \int dq\, w(q)\,(D(q)-2)^{2}$ is introduced heuristically. Strengthening the mathematical foundation, for instance by deriving these quantities from a variational principle or from established quantum gravity frameworks such as spin networks, path integrals, or spectral geometry, would elevate the proposal beyond phenomenology.

In this direction, a convenient way to motivate the appearance of the quantity $\gamma$ is to treat the
scale--dependent dimension $D(q)$ as a fluctuating field and determine its profile from
an information--theoretic variational principle.  Here $q$ denotes a resolution scale,
$D(q)$ the effective dimension, and $w(q)>0$ a weight encoding the density of modes.
The notation $P(D \,|\, q)$ denotes the probability of observing the effective dimension
$D$ when the geometry is probed at scale $q$.  To describe statistical fluctuations of
$D$, we introduce the Shannon entropy
\[
S=-\int dq\, w(q)\,P(D\mid q)\ln P(D\mid q),
\]
which follows from the standard definition of continuous entropy for a probability
distribution, weighted here by $w(q)$ to account for the number of modes contributing at
each scale.  Maximizing $S$ under the constraint that the quadratic deviation
$\langle(D-2)^2\rangle$ is fixed yields the Gaussian distribution
$P(D\mid q)\propto e^{-\lambda(D-2)^2}$ and an effective action
\[
I[D]=\lambda\int dq\, w(q)\,(D(q)-2)^2.
\]
Since the integrand contains no derivatives of $D$, the Euler--Lagrange equation gives
the extremum $D(q)=2$, and writing $D(q)=2+\delta D(q)$ shows that $I[D]$ is simply the
quadratic fluctuation action.  The on--shell value
\[
\gamma\equiv I[D]\sim\int dq\, w(q)\,(D(q)-2)^2
\]
thus measures the integrated variance of dimensional flow across scales and provides a
natural information--theoretic interpretation of $\gamma$ as the cost of deviating from
the UV fixed point $D=2$.

\section{Quantum dynamics from uncertainty minimization}

Quantum dynamics arises from minimizing an action consistent with the modified uncertainty relation, yielding a deformed kinetic operator. Let $\psi$ be a quantum state probing spacetime geometry at resolution $\ell \sim \hbar/\Delta p$. Then the induced position uncertainty satisfies
(\ref{eq1}) to leading order in $\ell_P$, where $\gamma$ is the geometric variance.
A convenient way to derive deformed quantum dynamics is to begin with this modified uncertainty relation (\ref{eq1}) or equivalent assuming the deformation of the canonical Heisenberg commutator (\ref{eq2}) which encodes minimal length, quantum-geometric constraints. To implement this at the operator level, one introduces a deformed momentum operator in position space $p \;\rightarrow\; \hat p = -i\hbar\,F(\partial_x)$, for instance $\hat p = -i\hbar\bigl(1-\gamma \, \ell_P^{2}\partial_x^{2}+\cdots\bigr)\partial_x$, ensuring that the modified commutation relation is satisfied. One then constructs a quantum action functional whose stationary points define the dynamics,
\[
\mathcal{S}[\psi]=\int dt\,dx\left[\frac{i\hbar}{2}\bigl(\psi^{*}\partial_t\psi-\psi\,\partial_t\psi^{*}\bigr)-\psi^{*}\hat H\,\psi\right],
\]
with the nonrelativistic Hamiltonian $\hat H=\hat p^2/(2m)+V(x)$. The kinetic operator is chosen so that wavefunctions saturate the modified uncertainty bound, which is equivalent to minimizing a generalized Fisher--information--type functional subject to normalization,
\[
\delta\!\left[\langle \hat p^{\,2}\rangle + \lambda\bigl(\langle x^{2}\rangle - \text{fixed}\bigr)\right]=0,
\]
where we are only allowed to vary the wavefunction among those that have a fixed second moment $\langle x^{2}\rangle$.
Solving this variational problem determines the structure of $\hat p^{\,2}$ and produces higher-order derivative corrections. The resulting deformed kinetic operator takes the form
\[
\hat H = -\frac{\hbar^{2}}{2m}\partial_x^{2} + \gamma\,\frac{\hbar^{2}\ell_{P}^{2}}{2m}\partial_x^{4}+ \cdots + V(x),
\]
and the corresponding effective Schr\"odinger equation becomes
\begin{equation}\label{Schr1}
i\hbar\,\partial_t\psi = \left[ -\frac{\hbar^{2}}{2m}\partial_{x}^{2} + \gamma\,\frac{\hbar^{2}\ell_{P}^{2}}{2m}\partial_{x}^{4} + V(x) \right]\psi.
\end{equation}
which is the emergent quantum dynamics consistent with the modified uncertainty principle. In such Schr\"odinger equation
the higher-derivative term $\partial_{x}^{4}$ encodes Planck-scale corrections to the kinetic energy, i.e., short-distance (microscopic) geometric effects, with nonlocality suppressed by $\ell_{P}^{2}$. Therefore, the uncertainty relations constrain admissible wavefunctions; minimizing uncertainty corresponds to extremizing an action; the extremum fixes a modified kinetic operator; and quantum dynamics arises as the Euler--Lagrange equation of this constrained variational principle. In summary, quantum dynamics is not postulated but emerges from an action whose kinetic term is determined by saturating a generalized uncertainty principle.

\smallskip

However, the claim that quantum mechanics is emergent sits uneasily with the use of $\hbar$ and standard uncertainty relations as inputs, risking circularity unless an $\hbar$-like quantity is itself derived from geometry.

A natural route to obtaining an $\hbar$--like quantity from geometry is to show that noncommutativity and uncertainty arise from coarse--graining fluctuating microscopic geometry, with the resulting commutator coefficient fixed by a geometric invariant. Consider a probe moving in a spacetime whose microstructure is encoded by a scale parameter $q$, an effective dimension $D(q)$, and metric fluctuations $g_{ab}(q,x)$,
so that the classical phase--space variables $(x,p)$ are emergent. The fundamental symplectic form $\omega = dp\wedge dx$ is then modified at small scales, $\omega_{\mathrm{eff}} = dp\wedge dx + \delta\omega$, where $\delta\omega$ arises from curvature fluctuations, dimensional flow, or horizon microstructure.  Coarse--graining over geometric microstates yields
\[
\langle \omega_{\mathrm{eff}} \rangle = (1+\hbar_{\mathrm{geom}})\, dp\wedge dx,
\]
with $\hbar_{\mathrm{geom}}$ the proportionality coefficient of the averaged correction.
Since quantization replaces Poisson brackets by commutators, the geometrically deformed
bracket $\{x,p\}_{\mathrm{eff}} = 1+\hbar_{\mathrm{geom}}$ leads to
\[
[x,p] = i\,\hbar_{\mathrm{geom}},
\]
so that the uncertainty relation $\Delta x\,\Delta p \ge \tfrac{1}{2}\hbar_{\mathrm{geom}}$
emerges statistically rather than being postulated.  If geometric fluctuations are
controlled by deviations from the UV fixed point $D=2$, then the symplectic defect is
set by the integrated variance of dimensional flow,
\[
\hbar_{\mathrm{geom}} \propto \int dq\, w(q)\,\bigl(D(q)-2\bigr)^{2},
\]
the same structure that defines $\gamma$.  Thus $\hbar_{\mathrm{geom}}$ acquires a clear
interpretation as a macroscopic imprint of microscopic geometric complexity: it measures
irreducible phase--space noise induced by fluctuating geometry and vanishes in the
classical limit $D(q)=2$.  This framework shows that quantum behavior can arise from
geometric fluctuations and statistical coarse--graining, with $\hbar$ generated rather
than assumed.

\smallskip

A concise way to relate $\hbar_{\mathrm{geom}}$ and $\gamma$ is to note that, as defined,
they are mathematically the same scalar functional, differing only by normalization and
physical interpretation.  Since both arise from the integrated variance of dimensional
flow,
\[
\gamma \sim \int dq\, w(q)\,(D(q)-2)^{2},
\qquad
\hbar_{\mathrm{geom}} \propto \int dq\, w(q)\,(D(q)-2)^{2},
\]
one has $\hbar_{\mathrm{geom}} = C\,\gamma$ for some constant $C$.  The distinction is
therefore conceptual: $\gamma$ is a dimensionless measure of geometric roughness,
capturing the strength of fluctuations in $D(q)$, whereas $\hbar_{\mathrm{geom}}$ plays
the role of an effective quantum of action, entering the commutator
$[x,p]=i\hbar_{\mathrm{geom}}$ and thus setting the scale of emergent quantum effects.
Because $\hbar$ carries units of action while $\gamma$ does not, one must introduce a
geometric scale $A_{0}$ with dimensions of action so that
$\hbar_{\mathrm{geom}} = A_{0}\,\gamma$.  Possible origins of $A_{0}$ include
Planck-scale geometry, horizon thermodynamics, or microscopic symplectic normalization.
Even if proportional, the two quantities live at different conceptual levels:
$\gamma$ characterizes geometry itself, while $\hbar_{\mathrm{geom}}$ characterizes the
dynamics emerging from that geometry.  They coincide only in natural units or if all
dimensional factors are absorbed into $w(q)$, but such a choice obscures the physics.
The nontrivial problem is explaining why $\hbar_{\mathrm{geom}}$ becomes universal and
constant, whereas $\gamma$ generically depends on scale and state; resolving this is
essential for recovering standard quantum mechanics.

\smallskip

A scale--dependent geometric quantity does not automatically yield a universal constant
$\hbar$; obtaining such universality requires additional structure that forces the
emergent coefficient to become Renormalization Group (RG)--invariant, self--averaging, and symmetry--protected.
Starting from
\[
\hbar_{\mathrm{geom}} \sim A_{0}\!\int dq\, w(q)\,(D(q)-2)^{2},
\]
one sees that, in general, $D(q)$ and $w(q)$ vary with scale, environment, and state,
implying $\hbar_{\mathrm{geom}}=\hbar_{\mathrm{geom}}(q,\text{state})$, whereas
experiment demands a constant $\hbar$.  A universal value can arise if geometric flow
approaches an RG fixed point so that $(D(q)-2)^{2}$ follows a universal crossover
function, making the integral RG--invariant; if spacetime microstructure is composed of
many patches whose fluctuations self--average so that only the ensemble variance
$\langle(D-2)^{2}\rangle$ survives; or if the coefficient of the commutator
$[x,p]=i\hbar_{\mathrm{geom}}$ is symmetry--protected by the consistency of the
Heisenberg algebra under canonical transformations.  A fundamental geometric scale
$A_{0}$ (from Planck geometry, horizon thermodynamics, or microscopic symplectic
normalization) must also fix the units of action.  Thus recovering a universal $\hbar$
requires: (i) universal geometric flow, (ii) a fixed measure $w(q)$, (iii) a fundamental
normalization scale $A_{0}$, and (iv) algebraic consistency across phase space.  When
these conditions hold, $\hbar$ emerges as an RG--invariant, statistically stable,
symmetry--enforced order parameter of spacetime microstructure; otherwise one obtains a
variable--$\hbar$ theory, which is experimentally excluded.

\subsection{From this Schr\"odinger equation to the modified semiclassical gravity}

To couple this quantum system to gravity semiclassically, we need the expectation value of the energy--momentum tensor. In the nonrelativistic regime, the dominant component is the energy density $T_{00}$, which is essentially the Hamiltonian density. For a wavefunction $\psi(x)$, the standard kinetic energy density is $\mathcal{E}_{\text{kin}}^{(0)}= \hbar^2/(2m)\,|\nabla\psi|^{2}$,
coming from the usual term $-\hbar^{2}\nabla^{2}/(2m)$. The GUP-induced correction modifies the kinetic operator by a factor that depends on $p^{2}$ (or equivalently on higher spatial derivatives). At leading order, this can be encoded as an effective correction
\[
\mathcal{E}_{\text{kin}}
\simeq
\frac{\hbar^{2}}{2m}\,|\nabla\psi|^{2}
\left[1 + \gamma\,\frac{\langle p^{2}\rangle}{m_{P}^{2}}\right],
\]
where $\langle p^{2}\rangle$ is the expectation value of the squared momentum in the state $\psi$. Thus, the expectation value of the energy density becomes
\[
\langle T_{00} \rangle= \frac{\hbar^{2}}{2m}\,\langle (\nabla\psi)^{2} \rangle \left[1 + \gamma\,\frac{\langle p^{2}\rangle}{m_{P}^{2}}\right]
+ \langle V(x)\rangle,
\]
where we have separated the potential contribution $\langle V(x)\rangle$ for clarity.
In semiclassical gravity, the spacetime geometry responds to the quantum expectation value of the energy--momentum tensor, then we have
\[
G_{\mu\nu} = 8\pi G\,\langle T_{\mu\nu} \rangle_{\text{geom}}.
\]
Here, $\langle T_{\mu\nu} \rangle_{\text{geom}}$ includes the above GUP-corrected energy density. The Planck-suppressed factor
$1 + \gamma\,\langle p^{2}\rangle/m_{P}^{2}$
then feeds into the right-hand side of Einstein's equations, leading to small but conceptually important corrections to the geometry. When applied to black hole backgrounds, these corrections manifest as entropy corrections, modified Hawking evaporation, and logarithmic terms that are known to be universal across many quantum gravity approaches.

\subsection{Emergent Schr\"odinger Dynamics from Geometric Noise}

Assume that near a rough/multifractal horizon geodesics fluctuate stochastically, that the metric fluctuations induce random accelerations,
and the noise is stationary and isotropic at small scales.
Let $dx^{i} = v^{i} dt + d\xi^{i}$, where $\langle d\xi^{i} d\xi^{j} \rangle = 2D\,\delta^{ij}\,dt$
with diffusion constant $D \sim \hbar/(2m)$. Then one obtains the Schr\"odinger equation
\[
i\hbar\,\partial_{t}\psi = \left(-\frac{\hbar^{2}}{2m}\nabla^{2}+V_{\mathrm{eff}}\right)\psi,
\]
where $V_{\mathrm{eff}}$ is the effective potential, a term that collects everything acting on $\psi$
beyond the free kinetic operator $\hbar^2 \nabla^{2}/(2m)$.
Now the Schr\"odinger dynamics arises as a hydrodynamic limit and he wavefunction encodes statistical geometry.
This mirrors Nelson's stochastic mechanics \cite{Ne1,Ne2}, the fractal paths of Feynman--Hibbs \cite{Fey,Sch}, and Nottale's scale relativity \cite{Not1,Not2}. Hence quantum mechanics emerges from geometric fluctuations.

\subsection{Quantization of fields on multifractal backgrounds}

We define the effective multifractal Laplacian by spectral averaging:
\[
\Delta_{\mathrm{MF}} := \int dq\, w(q)\, (-\Delta)^{D(q)/2},
\]
that is a self-adjoint pseudodifferential operator. In order to find the effective kinetic operator, for $D(q)=2+\Delta(q)$ with $|\Delta(q)| \ll 1$, we have $\Delta_{\mathrm{MF}}= -\Delta + \gamma\,\Delta \ln^2(-\Delta) + \cdots$, because
\[
(-\Delta)^{1+\Delta(q)/2} = -\Delta - \frac{\Delta(q)}{2}\,\Delta\ln(-\Delta) - \frac{\Delta^2(q)}{8}\,\Delta\ln^2(-\Delta) + \cdots,
\]
and averaging over $w(q)$ eliminates linear terms and produces the variance $\gamma \varpropto \langle \Delta^2(q) \rangle$.
Then, the dynamics of a scalar quantum field $\psi$ is governed by the multifractal Schr\"odinger equation
\[
i\hbar\,\partial_t \psi = \left[-\frac{\hbar^2}{2m}\left(\Delta - \gamma\,\Delta \ln^2(-\Delta)\right) + V\right]\psi.
\]

Moreover, the propagator behaves as $G(k) \sim 1/(k^2(1 + \gamma \ln^2(k^2)))$, which improves the UV behaviour of loop integrals; the theory is weakly nonlocal at scales $\lesssim \ell_P$, with effective locality restored in the IR.

\smallskip

Summarizing this section fractal geometry affects the quantum dynamics through GUP that now is geometric not phenomenological with a GUP parameter $\beta$ given by $\gamma = \mathrm{Var}[D(q)]$. Moreover, from the effective Schr\"odinger equation obtained we get the modified semiclassical gravity. Finally, quantum mechanics is emergent from geometric fluctuations, and not fundamental.

\section{Entropic Gravity Connection}

In this section we are going to see that gravity is obtained as thermodynamic backreaction and Einstein equations arise as an equation of state. Hence, gravitational dynamics emerge as the thermodynamic response of multifractal geometry to information/entropy flow $S \sim \alpha(\ell) A /(4 \ell_{P}^{2})$, with Einstein gravity recovered in the smooth limit.

\smallskip

The scale-independent entropy of a multifractal horizon have been determined in \cite{Gine} and results
\begin{equation}\label{entsi}
S = \frac{A}{4\ell_{P}^{2}} \int dq\, w(q)\,\ell_{P}^{\,2-D(q)} .
\end{equation}
Taking into account the Verlinde gravity theory \cite{Ver}, entropy gradient is given by
$F\,\delta x = T\,\delta S$, with Hawking temperature $T \sim 1/M$, that gives
\[
F = \frac{GMm}{r^{2}}\left[1 + O(\langle \Delta^{2}(q) \rangle)\right],
\]
where $\Delta(q)$ denotes a small fluctuation of the separation around the mean distance $r$, that is, $\Delta(q)$ represents microscopic, stochastic, or fractal deviations of the particle's trajectory (depending on the framework--stochastic mechanics, path-integral fluctuations, scale relativity, etc.). Moreover, $\langle \Delta^{2}(q) \rangle$
is the mean-square fluctuation (the variance) of that deviation. Consequently, the term
$O(\langle \Delta^{2} \rangle)$ indicates higher-order corrections to the classical Newtonian force arising from those fluctuations.
In brief, $\Delta(q)$ encodes small-scale non-classical deviations of the particle's position, and the formula says Newton's law is recovered up to corrections controlled by the size of those fluctuations.
Hence, Newtonian gravity emerges from statistical geometry, with corrections controlled by the multifractal spectrum.

\smallskip

Summarizing, gravity is the entropic response of spacetime microstructure, and quantum uncertainty can be interpreted as geometric roughness. The indeterminacy of quantum observables does not arise from measurement limitations alone, but from the intrinsically irregular, non-smooth structure of spacetime or particle trajectories at microscopic scales. In this view, quantum fluctuations reflect an underlying fractal or stochastic geometry, so that uncertainty emerges as a manifestation of unresolved geometric detail rather than fundamental randomness.
Then both describe the same underlying physics, revealed through different limiting regimes.

\smallskip

Hence, in the described framework we get a geometric derivation of the GUP, a statistical origin of non-commutativity, an emergent Schr\"odinger equation, a thermodynamic origin of gravity, an a unified language for entropy, uncertainty, and geometry.
Without the need to quantize the spacetime, introduce speculative degrees of freedom, without violating GR or QM.
Indeed, we have an effective quantum gravity framework, a geometric Effective Field Theory (EFT) of Planck-scale physics, and compatible with Loop Quantum Gravity (LQG), Asymptotic Safety, Causal Dynamical Triangulations (CDT), and relativistic statistical mechanics.

However, is not complete UV theory (in the sense that is not a fundamental theory that consistently describes physics at all energy scales), is not a replacement for canonical quantization, or a speculative metaphysics.

\smallskip

Quantum behavior, generalized uncertainty principles, and gravitational dynamics may all be understood as emergent consequences of microscopic spacetime geometry. In this view, quantum gravity is not obtained by quantizing geometry, but by recognizing that geometry itself behaves as a statistical system whose coarse-grained description reproduces quantum mechanics and gravity simultaneously.
Hence, Quantum mechanics, generalized uncertainty principles, and gravity can be understood as successive effective descriptions of a fundamentally multifractal spacetime geometry probed at different resolutions.

\section{Emergent gravity from multifractal thermodynamics}

Following Jacobson \cite{Jac}, consider an arbitrary spacetime point $p$ and a local causal horizon generated by a congruence of null geodesics \cite{HaEl,Wal} with tangent $k^{\mu}$, which we approximate as a local Rindler horizon for an accelerated observer. The observer sees an Unruh temperature $T = \kappa/(2\pi)$ \cite{Unr}, where $\kappa$ is the local surface gravity (acceleration scale). The Clausius relation $\delta Q = T\,dS$ is imposed for all such local Rindler horizons through $p$, with $\delta Q$ interpreted as the energy flux across the horizon and $dS$ the corresponding change in horizon entropy and this yields the Einstein equation as an equation of state \cite{Jac,Eli,Pad}.

The scale-dependent entropy of a multifractal horizon is $S(\ell) = \alpha(\ell) A /(4\ell_{P}^2)$, where $\ell$ is the characteristic linear size of the horizon. For a smooth 2-dimensional surface, $A \sim \ell^2$ and that relation is purely geometric and independent of quantum gravity. If we insert this relation in $(\ell/\ell_P)^{2-D(q)}$ we get $(A/\ell_P^2)^{(2-D(q))/2}$ and we arrive to the scale-dependent horizon entropy associated with an area element $A$ is given by
\[
S = \frac{A}{4\ell_P^2}
\int dq\, w(q)\, \left (\frac{A}{\ell_P^2} \right )^{\,\frac{2-D(q)}{2}}.
\]
For small deviations $D(q) = 2 + \Delta(q)$, one finds
\begin{equation}
S= \frac{A}{4\,\ell_{P}^{2}}
\left[1 - \frac{\langle \Delta(q) \rangle}{2} \ln \left (\frac{A}{\ell_{P}^2} \right)
+ \frac{\langle \Delta^{2}(q) \rangle}{8} \ln^2 \left (\frac{A}{\ell_{P}^2} \right)
+ \cdots \right],
\end{equation}
and different assumptions about $\langle \Delta(q) \rangle$ and $\langle \Delta^2 (q) \rangle$ correspond to different known QG models (LQG, string theory, asymptotic safety, etc.). If the mean dimensional shift does not vanish $\langle \Delta(q) \rangle \ne 0$ corresponds to a systematic shift of the effective horizon dimension. Inside this case we have the standard QG logarithmic correction \cite{Kaul} where we have
\[
S = \frac{A}{4\ell_P^2}+ c \ln \left (\frac{A}{\ell_{P}^2} \right)+ \cdots,
\]
where $c \varpropto \langle \Delta(q) \rangle$. Indeed $c=- A \langle \Delta(q) \rangle /(8 \ell_p^2)$ which depends on $A$, so it is not exactly the same as the usual constant as LQG/string/asymptotic safety formulas. Nevertheless to match the standard form with a constant $c$ one may fix a reference area
$A_0$ and define the coefficient $c$ at that scale, for instance by expanding the entropy around $A_0$.
A second case is when  $\langle \Delta (q) \rangle = 0$ and  $\langle \Delta^2(q) \rangle \ne 0$
\[
S = \frac{A}{4\ell_P^2}
\left[1 + \frac{\gamma}{8}\ln^2\!\left(\frac{A}{\ell_P^2}\right) + \cdots\right], \quad  \mbox{where} \quad \gamma \varpropto \langle \Delta^2(q) \rangle.
\]
This case corresponds to a fractal horizon with stochastic geometry. Finally we have the case mixed case when both a nonzero $\langle \Delta(q) \rangle \ne 0$ and $\langle \Delta^2(q) \rangle \ne 0$ that we have both contributions of $\ln(A/\ell_P^2)$ and  $\ln^2(A/\ell_P^2)$
that correspond to general multifractal models. In any case, the leading term reproduces the Bekenstein--Hawking area law \cite{Bek1,Bek2,Haw1,Haw2}, while the subleading term gives logarithmic corrections. In the second case, the variation of the entropy under a small change of the horizon area $\delta A$ is then
\[
\delta S = \frac{\delta A}{4\ell_P^{2}} \left[1 + \frac{\gamma}{4}\,\ln\!\left(\frac{A}{\ell_P^2}\right) + \frac{\gamma}{8}\,\ln^{2}\!\left(\frac{A}{\ell_P^2}\right) + \cdots \right],
\]
where the $\gamma$-dependent terms are Planck-suppressed corrections.

\subsection{From horizon thermodynamics to Einstein's equations}

In this section, following Jacobson's thermodynamic construction, we consider an arbitrary spacetime point and associate a local Rindler horizon. Matter crossing the horizon produces a heat flux determined by the energy--momentum tensor, while the corresponding geometric response is described by the focusing of the null generators through the Raychaudhuri equation. Equating both quantities through the Clausius relation leads to the Einstein field equations.

Hence, we consider an arbitrary spacetime point $p$ and construct a local Rindler horizon generated by a congruence of null geodesics with tangent vector $k^\mu$. Matter crossing this local causal horizon transports energy, which is perceived by the uniformly accelerated observer as a heat flux. This flux is determined by the local energy--momentum tensor according to
\[
\delta Q = \int_{\mathcal H} T_{\mu\nu}\, \chi^\mu\, d\Sigma^\nu,
\]
where $\chi^\mu$ is the approximate boost Killing vector associated with the local Rindler horizon and $d\Sigma^\nu$ is the horizon surface element.
Near the horizon,
\[
\chi^\mu\simeq-\kappa\lambda\,k^\mu,
\]
where $\kappa$ is the surface gravity and $\lambda$ is an affine parameter along the null generators. Therefore,
\[
\delta Q = -\kappa \int_{\mathcal H} \lambda\, T_{\mu\nu} k^\mu k^\nu\, d\lambda\,dA.
\]
This expression shows that the heat flux is completely determined by the null projection of the energy--momentum tensor.\vspace{0.1cm}

The energy crossing the horizon modifies its geometry through the focusing of the null generators. This geometric response is encoded in the expansion $\theta$ of the null congruence, whose evolution is governed by the Raychaudhuri equation,
\[
\frac{d\theta}{d\lambda} = -\frac12\theta^2 -\sigma_{\mu\nu}\sigma^{\mu\nu}- R_{\mu\nu}k^\mu k^\nu.
\]
Choosing the horizon to be instantaneously in local equilibrium at $p$, $\theta(0)=0$, $\sigma_{\mu\nu}(0)=0$,
the quadratic terms are negligible to first order in $\lambda$, yielding
\[
\theta(\lambda) \simeq - \lambda R_{\mu\nu} k^\mu k^\nu.
\]
Since the expansion measures the fractional change of the horizon area,
\[
\frac{dA}{d\lambda} = \theta A,
\]
the corresponding area variation becomes
\[
\delta A = - \int_{\mathcal H} R_{\mu\nu} k^\mu k^\nu \lambda\, d\lambda\,dA.
\]
Thus, while the heat flux is controlled by the matter content through $T_{\mu\nu}$, the change of entropy is governed by the spacetime curvature through $R_{\mu\nu}$.\vspace{0.1cm}

The bridge between matter and geometry is provided by the Clausius relation $\delta Q = T\,\delta S$, which is assumed to hold for every local Rindler horizon.
Using the geometric entropy derived in the previous section,
\[
\delta S = \frac{\delta A}{4\ell_P^{2}} \left[1 + \frac{\gamma}{4}\,\ln\!\left(\frac{A}{\ell_P^2}\right) + \frac{\gamma}{8}\,\ln^{2}\!\left(\frac{A}{\ell_P^2}\right) + \cdots \right],
\]
together with the Raychaudhuri relation for $\delta A$ and the expression for the heat flux, one finds that

\[
T_{\mu\nu} k^\mu k^\nu = \frac{1}{8\pi G} R_{\mu\nu} k^\mu k^\nu \left[ 1+O(\gamma) \right],
\]
where the $O(\gamma)$ terms originate from the multifractal correction to the horizon entropy.
Since this relation holds for every null vector $k^\mu$, it follows that
\[
R_{\mu\nu} - \frac{1}{2} R g_{\mu\nu} + \Lambda g_{\mu\nu} = 8\pi G \langle T_{\mu\nu}\rangle + O(\gamma),
\]
namely the Einstein field equations supplemented by multifractal corrections.
In the limit $\gamma\rightarrow0$, the standard Einstein equations are recovered as the thermodynamic equation of state of spacetime. The additional $O(\gamma)$ terms encode departures from classical geometry associated with the scale-dependent microstructure of spacetime and provide the origin of the modified horizon thermodynamics discussed throughout this work. In particular, these corrections are responsible for the modified horizon thermodynamics (entropy corrections, altered Hawking evaporation, etc.) that remain universal across many quantum gravity approaches.

\section{Conclusions}

In this work, from multifractal geometry and geometry fluctuations we have derived a multifractal GUP that allows to defines the quantum dynamics and the gravitational thermodynamics and finally arrive to an effective quantum gravity without invoking any quantum gravity theory.

In this perspective, quantum gravitational effects emerge primarily as statistical manifestations of the underlying microstructure of horizons, rather than as direct consequences of canonical quantization. This suggests that the apparent quantum behavior of spacetime may be deeply rooted in its geometric and statistical properties, reflecting collective effects of a complex, multifractal microgeometry. Although this approach does not yet constitute a complete theory of quantum gravity, it highlights how quantum-like phenomena could arise naturally from the structure of spacetime itself.
Building on both the current results \cite{Gine} and the findings presented here, one can envision an effective, semiclassical, and statistically grounded framework for quantum gravity. In this emergent picture, quantum gravitational features are encoded in geometric fluctuations and averaged microstructures, providing a coherent bridge between classical geometry and quantum behavior. While this does not yet provide a fundamental, fully quantized theory, it represents a substantial step toward a geometrically emergent understanding of quantum gravity, offering a concrete foundation for future developments in the field.\newline

The framework presented is conceptually well aligned with modern approaches to quantum gravity, particularly those emphasizing emergent spacetime, holographic or entropic gravity, and scale-dependent or fractal geometry. The central idea is that the generalized uncertainty principle and quantum behavior may arise from microscopic geometric fluctuations.

This work does not propose a replacement for Hilbert-space quantum mechanics, nor does it attempt to derive the Born rule or the measurement postulate. Instead, its scope is limited to suggesting that certain kinematical features usually regarded as quantum (specifically generalized uncertainty relations) may admit a geometric origin in the microscopic structure of spacetime. Standard quantum mechanics, including its probabilistic and measurement framework, is retained as the effective macroscopic description, while the geometric model is interpreted as a possible microscopic underpinning of quantum uncertainty. The proposal differs from effective field theory, which assumes quantum mechanics a priori, and from stochastic gravity, where fluctuations are externally prescribed; here the fluctuations arise intrinsically from spacetime geometry. It is also complementary to emergent-gravity approaches, focusing on the emergence of quantum behavior rather than gravitational dynamics. In short, the paper advances a limited and well-defined claim: that the generalized uncertainty principle may emerge from geometric microstructure, without asserting a full reconstruction of quantum mechanics.

The effective dimension $D(q)$ introduced in this work is defined in general terms as a scale-dependent geometric quantity characterizing microscopic spacetime structure, without assuming a specific quantum-gravity realization. Different approaches provide natural operational interpretations of $D(q)$. In Causal Dynamical Triangulations, a possible realization is the spectral dimension $D_{s}(\sigma)$, which may be related to the coarse-graining scale through $q\sim\sigma^{-1/2}$, allowing one to interpret $D(q)$ as $D_{s}(\sigma=q^{-2})$ as a consistent instance of dimensional flow. In Asymptotic Safety, $D(q)$ corresponds to the effective dimension along the renormalization-group trajectory, approaching the ultraviolet value \(D=2\). In Loop Quantum Gravity, although no running dimension is defined explicitly, \(D(q)\) may be viewed as an effective coarse-grained dimension arising from statistical fluctuations in spin-network punctures. Under these interpretations, the geometric parameter \(\gamma=\int dq\, w(q)\,[D(q)-2]^{2}\) measures the integrated variance of dimensional behavior across scales, providing a unified geometric characterization while keeping the formalism framework-independent.

A central feature of the present framework is that the generalized uncertainty principle does not rely on a universal deformation parameter, but instead on a geometric coefficient \(\gamma=\gamma[D(q)]\) arising from the scale-dependent dimensional flow of spacetime. This leads to several characteristic and potentially observable consequences. First, the effective GUP parameter becomes curvature dependent, since \(\beta=\beta[D(q)]\), implying that quantum corrections strengthen in regions where the effective dimension deviates from \(D=2\). Second, the logarithmic corrections to black-hole entropy need not be universal: the coefficient may acquire scale dependence through $\gamma$, producing modified evaporation and a running Hawking temperature. Third, gravitational-wave propagation and black-hole ringdown spectra may receive corrections controlled by the dimensional flow rather than by a fixed Planck-scale deformation, offering a possible observational probe of microscopic geometry. Additional consequences include scale-dependent modifications to dispersion relations, non-universal displacement noise in interferometry, and a running effective Planck length \(\ell_{\mathrm{eff}}=\ell_{P}[1+\gamma(q)]^{1/2}\). These effects arise directly from the geometric origin of the GUP and differ from scenarios in which the deformation parameter is fundamental and constant. Once a specific microscopic model determines the dimensional flow \(D(q)\) -- for example through CDT, Asymptotic Safety, or other approaches -- the framework predicts corresponding modifications to black-hole thermodynamics, high-energy propagation, and quantum uncertainty. In this sense, the geometric-emergence picture provides observational avenues for distinguishing itself from conventional GUP models while remaining agnostic about the detailed dynamics of \(D(q)\).

The geometric-emergence framework proposed here is not intended to supersede string theory, noncommutative geometry, or quantum-information approaches, but rather to offer a complementary perspective with distinct conceptual advantages. In contrast to models in which the deformation parameter of the generalized uncertainty principle is introduced axiomatically --such as the string-theoretic minimal length or the noncommutative deformation of the commutators-- the present approach interprets the GUP coefficient as a derived quantity, \(\gamma=\int dq\, w(q)\,[D(q)-2]^{2}\), determined by the scale-dependent geometric fluctuations of spacetime. This reduces the number of fundamental assumptions by attributing quantum corrections to the underlying dimensional flow rather than to postulated nonlocality or modified commutators. Moreover, the geometric picture naturally aligns with modern quantum-information ideas: the quantity \(\gamma\) may be viewed as an integrated measure of geometric information or complexity across scales, suggesting that microscopic geometry stores information and that quantum uncertainty reflects limited access to that information. In this sense, geometric emergence and quantum-information approaches need not be competing paradigms; geometry may constitute the microscopic substrate, while quantum information provides the effective language describing coarse-grained correlations within that substrate. Exploring a quantitative correspondence between dimensional flow, entanglement measures, and generalized uncertainty relations may therefore offer a promising bridge between geometric and information-theoretic descriptions of emergent spacetime.\vspace{0.1cm}

Overall, the central hypothesis explored in this work --that characteristic quantum behavior may emerge from the statistical properties of spacetime microgeometry-- constitutes a viable and conceptually promising research direction. Rather than treating generalized uncertainty relations or minimal length effects as fundamental postulates, the proposed framework seeks to interpret them as effective manifestations of an underlying geometric structure. Although the present analysis should be regarded as a proof of concept rather than a complete theory of quantum gravity, it suggests that geometric fluctuations and dimensional flow may provide a common language connecting several existing approaches, including Causal Dynamical Triangulations, Asymptotic Safety, and Loop Quantum Gravity. To evolve into a predictive and falsifiable theory, however, the framework requires a more rigorous mathematical foundation, including a derivation of its geometric functionals from a well-defined variational or statistical principle, a clearer ontological formulation specifying the status of quantum mechanics as an emergent effective description, explicit mappings onto established quantum-gravity formalisms, and quantitative predictions that distinguish it experimentally from other generalized uncertainty principle models. Particularly promising directions include the development of curvature-dependent generalized uncertainty relations, non-universal corrections to black-hole thermodynamics governed by dimensional flow, and possible observational signatures in gravitational-wave propagation, black-hole ringdown spectra, and high-precision interferometry. Furthermore, establishing connections between statistical microgeometry and quantum-information-theoretic concepts such as entanglement entropy and information flow may provide a deeper understanding of how spacetime geometry gives rise to effective quantum behavior. In this sense, the present work should be viewed as a first step toward a broader research programme aimed at deriving quantum phenomena from the statistical organization of spacetime itself.\newline


\noindent {\bf Competing Interests.} The author has no relevant financial or non-financial interests to disclose.\newline

\noindent {\bf Data Availability Statement.} Data sharing not applicable to this article as no datasets were generated or analysed during the current study.

\end{document}